\documentstyle[12pt]{article}

\newcommand{\mtx}[2]{\left(\begin{array}{#1}#2\end{array}\right)}

\begin{document}

\begin{center}

\bigskip
{\Large Entangled Rings}\\

\bigskip

Kevin M.~O'Connor and William K.~Wootters\\

\bigskip

{\small{\sl

Department of Physics, Williams College, Williamstown, 
MA 01267, USA }}\vspace{3cm}

\end{center}
\subsection*{\centering Abstract}
{Consider a ring of $N$ qubits in a translationally invariant
quantum state.  We ask to what extent each pair of nearest 
neighbors can be entangled.  Under certain 
assumptions about the form of the state, we find a formula
for the maximum possible nearest-neighbor entanglement.
We then compare this maximum with the entanglement achieved 
by the ground state of an antiferromagnetic ring consisting of 
an even number of spin-1/2
particles.  We find that,
though the antiferromagnetic ground state typically does not maximize
the nearest-neighbor entanglement relative to all other states,
it does so relative to other states having zero $z$-component of spin.}

\vfill

PACS numbers: 03.67.-a, 03.65.Bz, 75.10.J\vfill

\newpage

\section{Introduction: Entanglement Sharing}
Quantum entanglement, as exemplified by the singlet state of two spin-1/2 particles, $\frac{1}{\sqrt{2}}(\mid\uparrow \downarrow \rangle - \mid\downarrow
\uparrow \rangle )$, has been the subject of much study in recent years
 \cite{entanglement}, 
largely because of its connection with quantum 
communication \cite{communication} and
computation \cite{computation}.  Entanglement bears some resemblance to classical
correlation, but it differs in important respects, including the fact that
entangled objects can violate Bell's inequality \cite{Bell}.  Perhaps one of the most characteristic 
differences is this:  if two similar quantum objects are completely 
entangled with each other, then
neither of them can be at all entangled with any other object, 
whereas there is 
no such restriction on classical correlations.  This property
is sometimes called the ``monogamy'' of entanglement.  
For the special case of three binary
quantum objects---three qubits---a quantitative extension of this rule 
has been proven in terms of a measure of entanglement called 
the ``concurrence'' which takes values between zero and one:
the square of the concurrence between qubits A and B, plus the square of the
concurrence between qubits A and C, cannot exceed unity \cite{CKW}.  
In other words, to the
extent that qubits A and B are entangled with each other, they limit the
entanglement between qubits A and C.  

The present paper further explores the degree to which entanglement 
can be shared among a number of qubits.  We focus on two closely related
but distinct problems.
(i) We consider a ring of $N$ qubits in 
a translationally invariant pure quantum state and ask to what extent nearest
neighbors can be entangled with each other; specifically, we ask how large the
nearest-neighbor concurrence can be.  
Note that in this first problem there is no Hamiltonian
specified; we are simply asking about the 
entanglement characteristics of quantum states.
(ii) For our second problem we consider a particular physical system, namely
a ring of $N$ spin-1/2 particles interacting via the Heisenberg antiferromagnetic
Hamiltonian, and ask whether the ground state of this system is a state of
maximum nearest-neighbor entanglement.  We will find that the
antiferromagnetic ring maximizes entanglement within a limited set of
states, but not absolutely.  

In both of these problems, we are focusing on 
{\em pairwise} entanglement within a system of 
$N$ particles.
At least three problems with a similar focus have been considered before.
D\"ur \cite{Dur} has shown that given a system of $N$ qubits and any specified set
of pairs of those qubits, one can design
a state such that all the pairs in the chosen set are entangled and all 
the other pairs are not.   
Koashi {\em et al.} \cite{Koashi} have
studied completely symmetric states of $N$ qubits
and have found that the maximum possible 
concurrence between pairs
is exactly $2/N$.  Thus in this context where
all the qubits are required to be equally entangled with
each other, the pairwise entanglement goes to zero in 
the limit of an infinite collection.  
Wootters \cite{chain} has 
considered a different problem, in which the qubits are arranged in an infinite
line and only the nearest-neighbor entanglement is maximized.  He found
that for the infinite chain in a translationally invariant state, the 
nearest-neighbor concurrence does not
have to be zero but can be as large as 0.434.  It is not yet known whether this
value is optimal.  The problem we are about to address is the simplest
finite version of the infinite chain problem. 

There have been several other studies of entanglement in $N$-component
systems, usually focusing on higher-order rather than 
pairwise entanglement \cite{other}. 
All of these studies contribute to our understanding
of entanglement distributed among many objects.  We hope that our
present results can eventually be combined with other work 
to construct a 
general theory of entanglement-sharing, not 
limited to qubits or to any particular geometry.      

\section{Maximizing Nearest-Neighbor Entanglement}
We begin by recalling the definition of the 
concurrence \cite{Hill,proof} between a pair
of qubits, which we will think of as spin-1/2 particles.  
Let $\rho$ be the density matrix of the pair, expressed in the
standard basis \hbox{$\{\mid\uparrow \uparrow \rangle, 
\mid\uparrow \downarrow \rangle, 
\mid\downarrow \uparrow \rangle, 
\mid\downarrow \downarrow \rangle \}$.}  
Let $\tilde{\rho}$ be the spin-reversed density matrix, defined by
$\tilde{\rho} = (\sigma_y \otimes \sigma_y) \rho^T
(\sigma_y \otimes \sigma_y)$,
where $\sigma_y$ is the matrix $\mtx{cc}{0&-i\\
i&0}$ and the superscript $T$ indicates transposition.  Then the 
concurrence of $\rho$ is given by $C = {\rm max}\{\lambda_1-\lambda_2-\lambda_3-\lambda_4,0\}$, where the $\lambda_i$ are the square roots
of the eigenvalues of $\rho \tilde{\rho}$ in descending order.  (These eigenvalues
are guaranteed to be real and non-negative even though $\rho \tilde{\rho}$
is not necessarily Hermitian.)  Concurrence is justified as a measure of 
entanglement by a theorem \cite{proof} showing that 
$C$ is a monotonically increasing
function of the entanglement of formation \cite{formation}.  
As we mentioned above, the values of concurrence
range from zero, for an unentangled state, to one, for a completely 
entangled state such as the singlet state.  

We imagine a set of $N$ particles arranged in a ring, 
with the locations of the particles labeled by an 
integer $i = 1, \ldots, N$.  In defining our problem, we restrict our attention
to translationally invariant pure states $|\psi\rangle$, 
that is, states that under the cyclic permutation
$i \rightarrow i+k$ (mod $N$) are changed by at most an overall
phase factor.  This restriction forces the concurrence to be the same for
each pair of nearest neighbors.  The problem, then, is simply to find the 
maximum possible value of this 
concurrence.\footnote{For a general, non-translationally-invariant
state, one could define at least two distinct problems along similar lines: (i) maximize 
the {\em average} entanglement over all nearest-neighbor pairs, and (ii) maximize
the {\em minimum} entanglement of all nearest-neighbor pairs.  The first of these
problems could be sensitive to the measure of entanglement one is
using---{\em e.g.}, concurrence, squared concurrence (also called the ``tangle''), or
entanglement of formation---even though these are all monotonic functions
of each other.
Problem (ii), which does not have this sensitivity, may thus be more interesting;
it may also reduce to the translationally
invariant problem considered here.} 

We have not yet been able to solve this problem in full.  We solve 
instead
a more tractable problem in which we limit  
the set of states over which the maximization 
is to be done.  Specifically, we require our states to satisfy the following two
conditions:\footnote{Condition 2 breaks the symmetry 
between $\mid\uparrow\rangle$ and $\mid\downarrow\rangle$.  Our
choice to use $\mid\uparrow\rangle$ rather than 
$\mid\downarrow\rangle$ in the statement of 
this condition is arbitrary and
does not affect any of our conclusions.}
\begin{enumerate}
\item The state $|\psi\rangle$ of the ring is an eigenstate of 
the total $z$-component of spin.
\item Neighboring particles cannot both be in the state $\mid\uparrow \rangle$.
\end{enumerate}
{\noindent}Though we are clearly leaving out many possible states, 
it is plausible that the maximum value we obtain for our restricted 
problem will not be far from the absolute maximum.  
This is because our two 
conditions tend to favor 
states with high nearest-neighbor entanglement.
To see this, let us consider the density matrix $\rho$ of a pair of nearest neighbors,
obtained by tracing $|\psi\rangle\langle\psi|$ over 
all the other particles.
Condition 1 implies that for any pair of particles, there can be no
{\em coherent superposition} of basis states with different numbers of up-spins,
{\em e.g.}, $\mid\downarrow \downarrow\rangle$ and 
$\mid\downarrow \uparrow\rangle$,
because the corresponding states of the rest of the chain are
orthogonal.  The density matrix $\rho$ must therefore have the following
block diagonal form:\footnote{In fact translational invariance implies that the matrix elements 
$w$ and $x$ must be equal---the frequency of occurrence of 
$\mid\uparrow\downarrow\rangle$ in the ring must be the same as that
of $\mid\downarrow\uparrow\rangle$---but we will not need to use this
equality in what follows.}
\begin{equation}
\rho = \mtx{cccc}{v&0&0&0\\0&w&z&0\\0&\bar{z}&x&0\\0&0&0&y}. 
\label{rho1}
\end{equation}
One can show by direct calculation that the concurrence of this density
matrix is 
\begin{equation}
C = 2\,{\rm max}\{|z| - \sqrt{vy},0\}. \label{sqrt}
\end{equation}
Condition 2 implies that the matrix element $v$ is zero, so that the
neighboring-pair density matrix becomes 
\begin{equation}
\rho = \mtx{cccc}{0&0&0&0\\0&w&z&0\\0&\bar{z}&x&0\\0&0&0&y}
\label{goodform}
\end{equation}
and the concurrence becomes simply 
\begin{equation}
C = 2|z|.  
\end{equation}
Density matrices of the
form (\ref{goodform}) have been singled out in two recent studies as having
particularly high entanglement.  Specifically, Ishizaka and Hiroshima \cite{Ishizaka} 
have proven that such density matrices maximize entanglement
for a fixed set of eigenvalues when one of the eigenvalues is zero.
(They also show numerically that the form (\ref{rho1}) is optimal
when all four eigenvalues are non-zero.)  
Munro {\em et al.} \cite{Munro} have shown that certain states of the
form (\ref{goodform}) maximize concurrence for a fixed value of the 
purity, defined as Tr$(\rho^2)$.   
These studies suggest that our two conditions are consistent with
high entanglement, but they do not guarantee that we will be able
to reach the absolute maximum.  
Indeed, we will see below that for at least one
value of $N$, the optimal concurrence is {\em not} achievable
by any state satisfying our conditions.  Nevertheless, our solution
to the restricted problem will be useful in Section~3 where we
discuss antiferromagnetic rings, and it 
should also serve as a good starting point for future work
on the complete problem.

Condition 1 forces the ring's state $|\psi\rangle$ to have a fixed
number $p$ of up-spins and a fixed number $N-p$ of down spins, but it does 
not specify the value of $p$.  Our strategy will be to fix the values of both $N$ and
$p$ and to maximize the 
nearest-neighbor concurrence
over all states having these values and satisfying condition 2.  
This problem
turns out to be exactly soluble, so that one can write down an analytic
formula for the maximum concurrence $C_{max}(N,p)$.  We can then use this formula to find the
optimal number of up-spins, and hence the optimal concurrence, 
for any ring size $N$.

For fixed $N$ and $p$, the most general
state we are considering has the form
\begin{equation}
|\psi\rangle = \sum_{1\leq  i_1<\cdots <i_p \leq N}
        b_{i_1 \ldots i_p}|i_1 \ldots i_p \rangle ,  \label{psi}
\end{equation}
where $|i_1 \ldots i_p \rangle$ is the state in which the particles at
locations $i_1, \ldots, i_p$ have their spins up and all the other
particles have their spins down.   Though
the above sum requires values of $b$ only for sets
of indices that are in 
ascending order, for convenience we define $b$ to be
symmetric in all its indices and equal to zero if any two indices
have the same value.  The normalization condition on
the coefficients $b$ is
\begin{equation}
\sum_{1\leq  i_1<\cdots <i_p \leq N} |b_{i_1 \ldots i_p}|^2 = 1.
\end{equation}
The condition of translational invariance is expressed as
\begin{equation}
b_{i_1 \ldots i_p} = e^{ik\theta}b_{i_1+k \ldots i_p+k},
\end{equation}
where addition is understood to be mod $N$ and $e^{iN\theta}=1$.
Finally, in accordance with condition 2
above, the coefficients $b$ must satisfy the constraint
\begin{equation}
 b_{i_1 \ldots i_p}=0 \,\, \,{\rm if}\,\,\, i_n-i_m = 1 
\,\,\,{\rm for \,\, any} \,\,n,m = 1,\ldots,p.   \label{condition}
\end{equation}
That is,
no state is allowed in which two up-spins are adjacent.    

To find the concurrence between two neighboring particles, we need
to find the off-diagonal element $z$ of the two-particle density
matrix as expressed in Eq.~(\ref{goodform}).  Translational
invariance guarantees that the value of $z$ will
be the same for each pair of nearest neighbors; 
we consider a specific pair at locations $i$ and $i+1$. 
Taking the partial trace of $|\psi\rangle\langle\psi|$ over all the
other particles, we find that
\begin{equation}
z = \sum_{1\leq k_2 < \cdots < k_p \leq N} b_{i,k_2 \ldots  k_p}
\bar{b}_{i+1,k_2 \ldots k_p}, \label{zsumm}
\end{equation}
so that
\begin{equation}
C = 2|z| = \bigg|\sum_{1 \leq k_2 < 
\cdots < k_p \leq N}2 b_{i,k_2 \ldots  k_p}
\bar{b}_{i+1,k_2 \ldots k_p}\bigg|. \label{zsum}
\end{equation}
This form tells us immediately that the concurrence can be maximized
by choosing the coefficients $b$ to be real and non-negative: if we were to use 
complex values, then the concurrence 
could only be made larger, not smaller, by 
replacing each coefficient $b$ by its absolute value.  Let us therefore
restrict our attention to such real and non-negative states. 
In that case, translational invariance takes the simple form
\begin{equation}
b_{i_1 \ldots i_p} = b_{i_1+k, \ldots, i_p+k}.
\end{equation}
Thus, once the values of $b_{1,i_2 \ldots i_p}$ are fixed, 
all the other $b$'s are determined.  

The condition expressed by Eq.~(\ref{condition}), {\em i.e.}, that no
two up-spins should be adjacent, is an awkward one to enforce
directly.  It is therefore helpful to relate our problem to a 
simpler problem that does not have this constraint.  Roughly 
speaking, we do this by removing from the ring the 
site immediately to the right of each
up-spin.  More precisely, we consider a ring of 
$N-p$ particles with exactly $p$ up-spins, and we assign to every 
state $|\psi\rangle$ of our original ring 
(every state, that is, that satisfies our
conditions) a corresponding
state $|\phi\rangle$ of the smaller ring: 
\begin{equation}
|\phi\rangle = \sum_{1\leq j_1<\cdots <j_p \leq N-p}
        d_{j_1 \ldots j_p}|j_1 \ldots j_p \rangle .  \label{phi}
\end{equation}
The coefficients $d$ are defined in terms of the original
coefficients $b_{1,i_2\ldots i_p}$ with $1<i_2<\cdots <i_p$.  Let $j_2 = i_2-1,
j_3 = i_3-2, \ldots ,j_p = i_p-(p-1)$; then $d_{1,j_2 \ldots j_p}
\equiv \sqrt{N/(N-p)}b_{1,i_2 \ldots i_p}$.  The values of the 
other $d$'s are determined by translational 
invariance---that is, $d_{j_1 \ldots j_p} = d_{j_1+k, \ldots, j_p+k}$ 
(mod $N-p$)---and
as before, we take  
$d_{j_1 \ldots j_p}$ to be symmetric 
under permutations of the indices and equal to zero whenever
two indices have the same value.
The factor $\sqrt{N/(N-p)}$ is included in order to make
$|\phi\rangle$ normalized: translations around the ring generate 
fewer $d$'s than $b$'s, so that the $d$'s need to be larger.\footnote{For each collection 
of $b$'s that are equal because of translational invariance, there is
a corresponding set of $d$'s, and the ratio of the sizes of these sets
is always $N/(N-p)$.}
Let us define a pseudo-concurrence $C'$ of the smaller ring
by analogy with Eq.~(\ref{zsum}).
\begin{equation}
C' = \sum_{1\leq k_2 < \cdots < k_p \leq N-p} 2d_{j,k_2 \ldots  k_p}
\bar{d}_{j+1,k_2 \ldots k_p}, \label{C'}
\end{equation}
where we have omitted the absolute value sign since the
$d$'s are all real and non-negative.
Because our states of the small ring do not satisfy condition 2, $C'$
is not the nearest-neighbor concurrence of the state $|\phi\rangle$. 
However, because of the relationship between $d$ and $b$, 
we can use $C'$ to find the concurrence $C$ of our original ring:
\begin{equation}
C = \frac{N-p}{N} C'.  \label{CC'}
\end{equation}
Thus we want to find the maximum possible value of $C'$ over
all real and non-negative, translationally invariant states of 
the $(N-p)$-particle ring with exactly $p$ up-spins.  

To do this, let us 
rewrite Eq.~(\ref{C'}) in a more convenient form by introducing
the creation and annihilation operators
\begin{equation}
a_j^\dag = \mtx{cc}{0&1\\0&0} \hskip 1pc {\rm and} 
\hskip 1pc a_j = \mtx{cc}{0&0\\1&0},
\end{equation}
which act on the particle at location $j$ of the small ring 
and are expressed here in the basis 
$\{\mid\uparrow\rangle,\mid\downarrow\rangle\}$.  In terms of these operators,
Eq.~(\ref{C'}) becomes simply
\begin{equation}
C' = 2\langle \phi |a^{\dag}_{j+1}a_j |\phi\rangle .  \label{expectation}
\end{equation}
The value given by Eq.~(\ref{expectation}) is the same for all pairs
$\{j,j+1\}$.  We can therefore write $C'$ as the average of this
quantity over $j$:
\begin{equation}
C' = \frac{2}{N-p} \langle \phi |\sum_{j=1}^{N-p}
a^{\dag}_{j+1}a_j |\phi\rangle  . \label{C1}
\end{equation}
Again using our assumption that the coefficients are real, 
we can re-express Eq.~(\ref{C1}) as 
\begin{equation}
C' = - \bigg(\frac{1}{N-p}\bigg) \langle \phi | H |\phi\rangle ,  \label{EC}
\end{equation}
where 
\begin{equation}
H = - \sum_{j=1}^{N-p} (a^{\dag}_j a_{j+1} + a^{\dag}_{j+1}a_j ).
\end{equation}
In other words, a state $|\phi\rangle$ maximizes $C'$
if it minimizes the expectation value of the operator
$H$, as long as this minimum is achieved with 
only non-negative real values of the coefficients $d$.    

The operator $H$ is the Hamiltonian for the one-dimensional ferromagnetic $XY$ model; so our problem reduces to finding the lowest-energy state of this model with
exactly $p$ spins up.  This is a solved problem \cite{Lieb}.  The solution begins with the 
observation that the operators $a^\dag$ and $a$ are not quite fermionic 
creation and annihilation operators, since $[a_j,a_k] = [a_j,a^{\dag}_k] = 
[a^{\dag}_j,a^{\dag}_k] = 0$ for $j\neq k$, whereas truly fermionic
operators attached to different sites would anticommute.
It is helpful to define new creation and annihilation operators $c^{\dag}$
and $c$ that are genuinely fermionic:
\begin{equation}
c_j = {\rm exp}\, \Big[ i\pi \sum_{k=1}^{j-1} a^{\dag}_k a_k \Big] a_j;
\end{equation}
\begin{equation}
c^{\dag}_j = a^{\dag}_j
{\rm exp}\, \Big[ -i\pi \sum_{k=1}^{j-1} a^{\dag}_k a_k \Big] .
\end{equation}
In terms of the $c$ operators, we have
\begin{equation}
H = - \sum_{j=1}^{N-p} (c^{\dag}_j c_{j+1} + c^{\dag}_{j+1} c_j ) \,\,\,
{\rm if}\,\, p \,\, {\rm is \,\,odd}
\end{equation}
and
\begin{equation}
H = - \sum_{j=1}^{(N-p)-1} (c^{\dag}_j c_{j+1} + c^{\dag}_{j+1}c_j ) 
+ (c^{\dag}_{N-p} c_1 + c^{\dag}_1 c_{N-p}) \,
\,\, {\rm if}\,\, p \,\, {\rm is \,\,even}.
\end{equation}
For either odd or even $p$, the Hamiltonian can be diagonalized exactly, 
so that the system can be regarded as a collection of $p$ independent 
identical fermions.  For odd $p$, one finds that the energy eigenvalues of these 
fermions are $e_m = -2\cos\Big(\frac{2m\pi}{N-p}\Big)$, $m = 1, \ldots, N-p$,
whereas for even $p$ they are $e_m =-2\cos\Big(\frac{(2m+1)\pi}{N-p}\Big)$,
$m = 1, \ldots, N-p$.
The minimum value of $\langle\phi | H |\phi\rangle$ is 
the sum of the $p$ smallest values $e_m$, since in the ground
state the fermions will
occupy the $p$ lowest energy levels.  This sum turns out to be 
given by the following formula,
valid for both even and odd values of $p$.
\begin{equation}
E_{min} = -\frac{2\sin \Big(\frac{p\pi}{N-p}\Big)}{\sin \Big(\frac{\pi}{N-p}\Big)}.
\end{equation}

The state $|\phi\rangle$ corresponding to this energy
is the discrete version of the 
ground-state wavefunction of a set of $p$ hard beads on a loop of wire.
The coefficients $d_{j_1 \ldots j_p}$ associated with this state
can be taken to be real and 
non-negative, and the state is 
translationally invariant.  Thus the assumed conditions are met and we can 
use $E_{min}$ to find the maximum pseudo-concurrence $C_{max}'$ in 
accordance with Eq.~(\ref{EC}):
\begin{equation}
C_{max}' = -\frac{1}{N-p}E_{min} = 
\frac{ 2\sin \Big(\frac{p\pi}{N-p}\Big)}{(N-p)\sin \Big(\frac{\pi}{N-p}\Big)}.
\end{equation}
Finally, using the relation (\ref{CC'}), we get the maximum nearest-neighbor 
concurrence of our original ring of $N$ particles:
\begin{equation}
C_{max}(N,p) = 
\frac{2 \sin \Big(\frac{p\pi}{N-p}\Big)}{N\sin \Big(\frac{\pi}{N-p}\Big)}.
\label{C(N,p)}
\end{equation}
Again, this is the maximum value under the following assumptions:
(i) the ring has exactly $p$ spins up, and (ii) no two up-spins are adjacent.

For a given value of $N$, we now need to find out what value of $p$
maximizes $C_{max}(N,p)$.  For any fixed $N$ it is easy enough to 
carry out this maximization explicitly.  Consider, for example, the 
case $N=7$.  In a ring of seven particles, the number $p$ of up-spins can 
have any of the following values without violating our condition 2: $p =
0, 1, 2$, and 3.  Inserting these numbers into Eq.~(\ref{C(N,p)}) we get
the corresponding values of the concurrence: $C = 0$, 0.286,
0.462, and 0.286.  Thus for a ring of seven particles it is best 
(under our assumptions) to have 
two spins up and five spins down.  
We have carried out this sort of direct maximization for 
the first several values
of the ring size $N$, with the following results:

\bigskip
{\noindent}
\begin{tabular}{|c|ccccccccc|}\hline
$N$ & 2&3&4&5&6&7&8&9&10 \cr
\hline
$p_{opt}$ & 1&1&1&1 or 2&2&2&2&3&3 \cr
\hline
$C_{max}$ & 1.000 & 0.667 & 0.500 & 0.400 & 0.471 & 0.462 & 0.433 & 0.444 & 0.449\cr
\hline
\end{tabular}

\bigskip

{\noindent}Note that though the maximum concurrence tends to decrease 
with increasing $N$, it is by no means monotonic.  

It is interesting to find the limiting value of $C_{max}$ as $N$ goes to infinity.
To do this, we write Eq.~(\ref{C(N,p)}) in terms of $N$ and $\alpha \equiv p/N$, 
and hold $\alpha$ fixed as $N$ goes to infinity.  The result is
\begin{equation}
C_{max}(\alpha) = \frac{2}{\pi}(1-\alpha)
\sin\bigg(\frac{\alpha\pi}{1-\alpha}\bigg).
\label{Clim}
\end{equation}
This equation gives the maximum nearest-neighbor concurrence (under our
assumptions) for an infinite chain of spin-1/2 particles in which the overall density
of up-spins is $\alpha$.  It is reassuring that this formula is identical to
the one obtained in Ref.  \cite{chain}, which considered only infinite
chains.  Differentiating Eq.~(\ref{Clim}), one finds that the optimal value
of $\alpha$ is 0.300844, for which $C_{max} = 0.434467$.  This number
is thus our candidate for the maximum nearest-neighbor concurrence of
an infinite chain of qubits (as in Ref.  \cite{chain}).  Note that, perhaps
surprisingly, for rings of 5 and 8 particles, the maximum values of $C$ 
as given in the above table
are {\em smaller} than the limiting value
for an infinite chain.  This is no doubt because in these cases one is
near the ``borderline'' between two different values of $p_{opt}$, and 
neither is particularly good.  This fact also suggests that the cases $N=5$
and $N=8$ are the best places to look for examples in which the maximum
concurrence is {\em not} achieved by a state satisfying our 
conditions.    

Indeed, by relaxing condition 2, one {\em can} achieve higher entanglement
for $N=5$.  The state
\begin{equation}
|\psi\rangle = \frac{1}{\sqrt{5}}\left\{\sin\theta\Big[\mid\uparrow\uparrow
\downarrow\downarrow\downarrow\rangle + \cdots \Big]
+ \cos\theta\Big[\mid\uparrow\downarrow\uparrow\downarrow
\downarrow\rangle + \cdots \Big]\right\},
\end{equation}
where the ellipses stand for all translations of the given basis state,
has a nearest-neighbor concurrence $C = 0.468$ when $\theta = 0.302$,
which is better than the value shown in the above table.  We have looked
for similar numerical improvements for $N=6$, 7, 8, 9, and 10, in each
case relaxing condition 2 but preserving condition 1, and we have found
none (not even for $N=8$).    
Thus it is conceivable that our formula gives
the true maximum for certain values of $N$, though it does not
do so for all values.  In any case, it gives us a lower bound on
the maximum nearest-neighbor concurrence, which we will be
able to use in the following Section.

To close this section, we write down explicitly the neighboring-pair
density matrix for our optimal state of the infinite chain.
In the form (\ref{goodform}), the matrix elements $w$ and $x$
must both be equal to $\alpha$, the density of up-spins.  This
is because every up-spin is isolated, so that the probability of the pair state $\mid\uparrow\downarrow\rangle$ is the same as the probability
that the first particle has its spin up, and similarly for the probability
of $\mid\downarrow\uparrow\rangle$.  We already have the value of $z$,
namely, half the concurrence; so the density matrix is
\begin{equation}
\rho = \mtx{cccc}{0&0&0&0\\0&0.301&0.217&0\\0&0.217&0.301&0\\
0&0&0&0.398}.  \label{rhomax}
\end{equation}
This matrix is not quite one of the special states identified by 
Munro {\em et al.} \cite{Munro},
which maximize entanglement for a fixed purity of the density
matrix.  Such a state would have all three of the non-zero diagonal elements
equal to $1/3$.  The fact that it is not the same shows that our problem is 
not equivalent to the fixed-purity problem.  Nevertheless, it is 
interesting that the two results are as similar as they are.  

\section{Entanglement in an Antiferromagnetic Ring}
Though we introduced an effective Hamiltonian in order
to solve the preceding problem, the problem itself did not
specify any Hamiltonian.  We now consider a 
more concrete physical model of a ring of $N$ qubits, namely, an 
antiferromagnetic ring of spin-1/2 particles in which 
neighboring particles interact
via the Heisenberg Hamiltonian
\begin{equation}
H = \sum_i^N \vec{\sigma}_i\cdot \vec{\sigma}_{i+1}.
\end{equation}
Here $\vec{\sigma} = (\sigma_x,\sigma_y,\sigma_z)$ is 
the vector of Pauli matrices and, as before, the sum $i+1$ is
understood to wrap around to 1 when $i=N$.
This model has been studied extensively over many decades, much
of the foundational work having been done by Bethe in the 
early days of quantum mechanics \cite{Bethe}.  
In the spirit of Section 2 we ask a new 
question about the model: does
the ground state maximize the nearest-neighbor entanglement?
We restrict our attention to rings with an {\em even} number
of particles,
partly because the calculation is considerably simpler in that
case, and partly because the symmetry of the even-$N$ ground
state suggests an interesting refinement of our question, 
as we will see shortly.

For the antiferromagnetic ring there is good
reason to  
expect a connection
between minimizing the energy and maximizing the
entanglement.  Contrary to what one would expect classically,
the ground state is not simply the alternating state $\mid\uparrow\downarrow
\uparrow\downarrow\cdots\rangle$.  Though this alternating state
minimizes the energy due to the $\sigma_{z}$ part of the Hamiltonian, 
it does
not do so well for the $\sigma_{x}$ and $\sigma_{y}$ parts.  
By contrast, the ground 
state for $N=2$, which is the singlet state
\begin{equation}
|\psi\rangle = \frac{1}{\sqrt{2}}(\mid\uparrow\downarrow\rangle
-\mid\downarrow\uparrow\rangle),
\end{equation}
treats all directions of space equivalently since it is rotationally
invariant.  Intuitively, one expects that for a ring of $N$ particles,
each pair of nearest neighbors is ``trying'' to be in the singlet state 
in order to minimize
its own energy but is thwarted to some extent by the similar efforts 
of neighboring pairs.
Now, the singlet state is maximally entangled; so in a certain sense each pair of 
nearest neighbors, by trying to minimize its energy, is also trying to be entangled.  
We want to see whether the pairs go as far in this
direction as they possibly could, that is, whether they in fact maximize
the nearest-neighbor entanglement.  Though we do not yet know the 
maximum possible
value of this entanglement (because of the extra conditions we 
imposed on our states in Section 2), we can nevertheless 
use the result of Section 2 as a 
benchmark for evaluating the entanglement of the antiferromagnetic
ring.  For example, if the nearest-neighbor concurrence of the infinite chain
is less than 0.434467, we know
that the entanglement is not maximal. 

We begin by invoking some basic facts about the ground state of
an antiferromagnetic ring with an even number of particles \cite{Orbach}: it
is translationally invariant, and it is an eigenstate of the total
$z$-component of spin with eigenvalue zero.  
These properties guarantee that the density matrix 
of each pair of neighboring 
particles has the form
\begin{equation}
\rho = \mtx{cccc}{v&0&0&0\\0&w&z&0\\0&\bar{z}&w&0\\0&0&0&v}.
\label{balanced}
\end{equation}
Let $E$ be the ground-state energy of the system, so that $E/N$ is
the contribution from a single pair:
$E/N = {\rm Tr}\,[\rho (\vec{\sigma}\cdot\vec{\sigma})]$.  
We now re-express the energy
$E/N$ in terms of the matrix elements of $\rho$.  The matrix
$\vec{\sigma}\cdot\vec{\sigma}$, written explicitly in the standard basis,
is
\begin{equation}
\vec{\sigma}\cdot\vec{\sigma} = 
\mtx{cccc}{1&0&0&0\\0&-1&2&0\\0&2&-1&0\\0&0&0&1}.
\end{equation}
Thus 
\begin{equation}
E/N = {\rm Tr}\,[\rho (\vec{\sigma}\cdot\vec{\sigma})]
= 2(v-w+2\,{\rm Re}\,z) = 4(v+\,{\rm Re}\,z)-1,  \label{E}
\end{equation}
where we have used the fact that Tr\,$\rho = 1$.

It is useful at this point to write the matrix elements $v$ and $z$
in terms of the coefficients that define the ring's state $|\psi\rangle$.
Just as in Section 2, we can write $|\psi\rangle$ as
\begin{equation}
|\psi\rangle = \sum_{1\leq i_1<\cdots <i_p \leq N}
        b_{i_1 \ldots i_p}|i_1 \ldots i_p \rangle ,  \label{psi3}
\end{equation}
where $p$ now has the specific value $N/2$.  And just as before,
we have
\begin{equation}
z = \sum_{1\leq k_2 < \cdots < k_p \leq N} b_{i,k_2 \ldots  k_p}
\bar{b}_{i+1,k_2 \ldots k_p}.  \label{zsummm}
\end{equation}
The corresponding expression for the matrix element $v$ is
\begin{equation}
v = \sum_{1\leq k_3 < \cdots < k_p \leq N}
|b_{i,i+1,k_3 \ldots k_p}|^2.
\end{equation}
Note that changes in the phases of the coefficients $b$ do not 
affect $v$, though they do affect $z$.  In order to minimize the
energy as given in Eq.~(\ref{E}), we want to choose these phases
so that Re\,$z$ is as negative as possible.  For a fixed set of absolute
values of the $b$'s, this can be done be letting all the $b$'s be
real, with alternating signs given by
\begin{equation}
{\rm sign \,\, of}\,\,b_{i_1 \ldots i_p}=(-1)^{i_1+\cdots +i_p}, \label{sign}
\end{equation}
in which case every term of Eq.~(\ref{zsummm}) is negative or zero.
Thus for the ground state of this system, we can write the energy per particle as 
\begin{equation}
E/N = 4(v - |z|) - 1.  \label{Eabs}
\end{equation}
Now, recall that the concurrence of a density matrix of the
form (\ref{balanced}) is [Eq.~(\ref{sqrt})]
\begin{equation}
C =\,\, {\rm max}\,\{2(|z| - v),0\}.      \label{Cv}
\end{equation}
We thus arrive at the following expression for the concurrence 
$C_{gs}$ of the ground state of this system, assuming (as is the case) that the
ground-state energy is sufficiently negative to make $C_{gs}$ positive.
\begin{equation}
C_{gs} = -\frac{1}{2}[(E/N)+1].  \label{EtoC}
\end{equation}
This simple relationship depends on the fact that the
number of particles in the ring is even.  If $N$ were odd, 
the pair density matrix would not have the form (\ref{balanced})
and its concurrence would most likely not be a simple function
of the energy alone.    

The ground state energies of antiferromagnetic rings have been
computed for many values of $N$ \cite{Orbach,Bonner}, including the limiting case
$N \rightarrow \infty$ \cite{Hulthen}.  From these
results and Eq.~(\ref{EtoC}) we can immediately write down
the concurrences.  The following table shows the values
of $C_{gs}$ for several values of $N$, along with corresponding values of
$C_{max}$ that we computed in Section 2.  
The figure 0.386 appearing in
the table as the concurrence of the ground state of the infinite chain can
be written exactly as $2\ln 2 -1$.
 
\bigskip
\hskip 2pc
\begin{tabular}{|c|ccccccc|}\hline
$N$ & 2&4&6&8&10&$\cdots$ &$\infty$ \cr
\hline
$-E/N$ & 3.000&2.000&1.868&1.825&1.806&$\cdots$ & 1.773 \cr
\hline
$C_{gs}$ & 1.000 & 0.500 & 0.434 & 0.412 & 0.403 &$\cdots$ & 0.386\cr
\hline
$C_{max}$ &1.000& 0.500 & 0.471 & 0.433 & 0.449 & $\cdots$ & 0.434 \cr
\hline
\end{tabular}

\bigskip

{\noindent}Thus, though for very small rings the antiferromagnetic
ground states are as entangled as the states we found in Section 2,
for larger rings they fall short.  We can therefore conclude that
the ground state of an antiferromagnetic ring does not in general
maximize nearest-neighbor entanglement.  

There is, however, a more limited sense in which these ground states do
maximize entanglement; this is the refinement we mentioned earlier. 
Let us restrict our attention to the set of states
which, like the ground state, are translationally invariant and have zero
total $z$-component of spin.  We will call such states ``balanced.''
We now show that the antiferromagnetic
ground state maximizes entanglement within the set of balanced states.  

Let us divide the set of all balanced states into equivalence
classes, two states being called equivalent if their coefficients
$b_{i_1\ldots i_p}$ in Eq.~(\ref{psi3}) agree in magnitude, differing
only in their phases.  Of all the states in a given 
equivalence class, none has a greater nearest-neighbor concurrence than
the unique state in that class for which the phases are given 
by Eq.~(\ref{sign}).
This is because, as in the case of the ground state, such phases allow perfect 
constructive interference
in Eq.~(\ref{zsummm}).  To put it in
symbols, $C(\psi) \leq C(\psi_0)$, where $|\psi\rangle$ is a 
general balanced state and $|\psi_0\rangle$ is the state obtained
from $|\psi\rangle$ by adjusting the phases of the $b$'s in 
accordance with Eq.~(\ref{sign}).
Now, for $|\psi_0\rangle$, the
expectation value of the energy $\langle\psi_0 | H|\psi_0\rangle$
is given by the same expression as in Eq.~(\ref{Eabs}):
\begin{equation}
\frac{1}{N}\langle\psi_0 | H|\psi_0\rangle = 4(v - |z|) - 1. 
\end{equation}
The concurrence of $|\psi_0\rangle$ is given by Eq.~(\ref{Cv});
so we have
\begin{eqnarray}
C(\psi) \leq C(\psi_0) &= \,
{\rm max}\,  {\nonumber}
\left\{-\frac{1}{2}\left[\frac{1}{N}
\langle\psi_0 | H|\psi_0\rangle +1\right],0\right\}\\
&
\leq \,{\rm max}\,\{-\frac{1}{2}[(E/N)+1],0\}= C_{gs} .  \label{string}
\end{eqnarray}
The last inequality comes from the fact that the ground state 
minimizes the expectation value of the energy.  We
have thus shown that no balanced state has a nearest-neighbor 
concurrence larger
than that of the ground state.

For comparison with Eq.~(\ref{rhomax}), it is interesting to write
down explicitly the neighboring-pair density matrix for the ground 
state of an infinite 
antiferromagnetic chain.  This density matrix is uniquely determined by the value 
of the concurrence, $C = 2\ln 2 -1$, and by the fact that the state is 
rotationally invariant (the latter condition implies that $|z|+v = w$).  
One finds that 
\begin{equation}
\rho = \mtx{cccc}{0.102&0&0&0\\0&0.398&-0.295&0\\
0&-0.295&0.398&0\\0&0&0&0.102}.
\end{equation}
If we think of the spins of the antiferromagnetic chain as ``trying'' to
maximize their entanglement, then evidently they are using a rather
different strategy than the one we used in Section 2.  There is no longer
any prohibition against neighboring up-spins.  Indeed the presence 
of such up-spin pairs in the antiferromagnetic chain allows the off-diagonal
element $z$ to have a larger magnitude than in Eq.~(\ref{rhomax}), 
which is good for entanglement.
On the other hand, the presence of such pairs also forces the matrix element 
$v = \langle\uparrow\uparrow\mid
\hskip -2pt \rho \hskip -2pt \mid\uparrow\uparrow\rangle$
to be non-zero, which is what reduces the concurrence to   
a value less than our best value of Section 2.  

\section{Conclusions}
We have obtained two main results.  

First, for a ring of $N$ qubits in 
a translationally invariant state, we have found values of the
nearest-neighbor concurrences that we know to be achievable and that
for some values of $N$ may be optimal.  
At the least, they are lower bounds on the maximum possible
concurrences.  

Second, we have found that the ground state of an antiferromagnetic
ring with an even number of particles typically does not maximize the 
nearest-neighbor concurrence over all states,
but that it does achieve such a maximum over the set of 
translationally invariant states with
no net spin in the $z$ direction.  This set of ``balanced'' states
includes all the eigenstates of total spin with eigenvalue zero; so we 
can also say that the ground state maximizes $C$ relative to all 
the spin-0, or rotationally invariant states. 

Putting Sections 2 and 3 together, we can conclude that whatever
the maximum-concurrence states may be, they are certainly not
balanced.  In other words, for maximizing concurrence it is
best to have
one direction of spin favored over the opposite direction. 
This is perhaps counterintuitive, since a maximally unbalanced
state such as $\mid\uparrow\uparrow\uparrow\cdots\rangle$ is not
entangled at all.  

The subject of Section 3 represents an unusual mix: one does not often
associate entanglement with energy-minimization.  One might wonder
whether the entanglement-maximization property of antiferromagnetic
rings, limited though it is, is a special case of a more general connection
between energy and entanglement.  Do physical systems tend to favor 
entangled states over unentangled states?  In a straightforward interpretation
of this question, the answer would seem to be no.   {\em Ferromagnetic}
systems, for example, have ground states in which 
the spins are completely unentangled.  Perhaps one could identify
a special class of Hamiltonians with interesting entanglement-maximizing
properties, but at present it is not clear how large such a class might be.  

We would like to thank Daniel Aalberts for a number of very helpful
discussions.


\newpage

\begin{thebibliography}{99}

\bibitem{entanglement} See, for example, C.~H.~Bennett, H.~J.~Bernstein,   
      S.~Popescu, 
      and B.~Schumacher, 
       {Phys.~Rev.~A} {\bf 53}, 2046 (1996); V.~Vedral, M.~B.~Plenio, 
      M.~A.~Rippin, and P.~L.~Knight,
       {Phys. Rev. Lett.} {\bf 78}, 2275 (1997);
      M. Horodecki, P. Horodecki, and R. Horodecki, Phys. Lett. A {\bf 223}, 1 (1996).
\bibitem{communication} B. Schumacher, Phys. Rev. A {\bf 54}, 2614 (1996).
\bibitem{computation} D.~P.~DiVincenzo, { Science} {\bf 270}, 255 (1995).
\bibitem{Bell} J. S. Bell, Physics {\bf 1}, 195 (1964).
\bibitem{CKW} V. Coffman, J. Kundu, and W. K. Wootters, Phys. Rev. A {\bf 61},
052306 (2000); quant-ph/9907047.  See also D.~Bru\ss , 
Phys. Rev. A {\bf 60}, 4344 (1999).
\bibitem{Dur} W. D\"ur, quant-ph/0006105.
\bibitem{Koashi} M. Koashi, V. Bu\v{z}ek, and N. Imoto, quant-ph/0007086.
\bibitem{chain} W. K. Wootters, quant-ph/0001114.
\bibitem{other} See, for example, A. V. Thapliyal, Phys. Rev. A {\bf 59}, 3336
(1999); J. Kempe, Phys. Rev. A {\bf 60}, 910 (1999);
D. Aharonov, quant-ph/9910081;
C. H. Bennett, S. Popescu, D. Rohrlich, J. A. Smolin and A. V. Thapliyal,
quant-ph/9908073;
N. Linden, S. Popescu, B. Schumacher and M. Westmoreland,
quant-ph/9912039;
G. Vidal, W. D\"ur and J. I. Cirac, Phys. Rev. Lett. {\bf 85}, 658 (2000).
\bibitem{Hill} S.~Hill and W.~K.~Wootters, {Phys. Rev. Lett.} 
{\bf 78}, 5022 (1997).
\bibitem{proof} W.~K.~Wootters, {Phys. Rev. Lett.} {\bf 80}, 
2245 (1998).
\bibitem{formation} C.~H.~Bennett, D.~P.~DiVincenzo, J.~Smolin, 
and W.~K.~Wootters, 
       {Phys.~Rev.~A} {\bf 54}, 3824 (1996).
\bibitem{Ishizaka} S. Ishizaka and T. Hiroshima, Phys. Rev. A {\bf 62},
22310 (2000).
\bibitem{Munro} W. J. Munro, D. F. V. James, A. G. White, and P. G. Kwiat,
unpublished.
\bibitem{Lieb} E. Lieb, T. Schultz, and D. Mattis, Annals of Physics {\bf 16},
407 (1961).
\bibitem{Bethe} H. A. Bethe, Z. Physik {\bf 71}, 205 (1931).
\bibitem{Orbach} R. L. Orbach, Phys. Rev. {\bf 115}, 1181 (1959).
\bibitem{Bonner} J. C. Bonner and M. E. Fisher, Phys. Rev. {\bf 135}, A640 (1964).
\bibitem{Hulthen} L. Hulth\'en, Arkiv. Mat. Astron. Fysik {\bf 26A}, No. 1 (1938).

\end{thebibliography}
\end{document}